# High efficient 120W 1018nm single-frequency narrow linewidth amplification based on wide-tunable DBR fiber seed source

Pan Li[1,3*], Linfeng Li[1], Mingze Wang[1], KaiMing Cao[1], Ruihong Gao[1,3], Heshan Liu[1,3], Meng Shi[2,3], Ziren Luo[1,3*]

1. National Microgravity Laboratory, Institute of Mechanics, Chinese Academy of Sciences, Beijing 100190, China;

2. Key Laboratory of Space Utilization, Technology and Engineering Center for Space Utilization, Chinese Academy of Sciences, Beijing 100094, China;

3. University of Chinese Academy of Sciences, Beijing 100049, China;

**Abstract:** This paper reports the achievement of 120W single-frequency narrow linewidth 1018nm laser based on wide-tunable DBR fiber seed source. The DBR structure seed source uses 8mm long doped optical fibers with a line width of 3.25k. The wavelength tuning range of this seed source exceeds 1.5 nm with the temperature range from 1°C to 95°C. The tuning wavelength and temperature show extremely high linearity, and there is no mode hopping during the tuning process. By adopting a multi-level fiber amplification structure, selecting appropriate doped fibers and optimizing their length, an output power exceeding 120W of 1018nm laser has been achieved. Measurement results indicate that the slope efficiency of the main amplification 77.3%, with an amplified spontaneous emission (ASE) suppression ratio greater than 60 dB. he output linewidth is 10.3 kHz, and the beam quality factor $M^2$ is less than 1.3.

# 1 Introduction

Recently, Rydberg-atom-based microwave measurements which can achieve high sensitivity and wideband microwave signal detection attracted many attentions [1-5]. Generally, there are three typical methods for preparing Rydberg atoms: single-photon excitation[6], two-photon excitation[7], and multi-photon cascade excitation[8]. The method of single-photon excitation can directly excite ground-state atoms to Rydberg P states in a single step, but it results in a low transition intensity and requires the excitation photons predominantly be in the violet and ultraviolet wavelength ranges, which makes the process complicated. The multi-photon cascade excitation method has more intermediate states. It typically requires multiple laser sources from different wavelength bands. Compared to the single-photon and multi-photon excitation, the two-photon excitation method (such as using 852nm and 509nm two-photon excitation for cesium atoms) for achieving the preparation of Rydberg states offers two advantages. Firstly, the 509nm laser source is relatively easier to be obtained. Secondly, it can prepare Rydberg D-states or S-states, and provide a broader range of Rydberg energy levels that can be coupled with microwaves.

With the further expansion of the application of Rydberg atomic microwave measurement [9-11], it requires more higher output power of a 509nm single-frequency laser source, with power demands reaching several tens to hundreds of watts [12]. There are two commonly used methods to obtain a high power 509nm single-frequency green laser. One is to directly use a semiconductor narrow linewidth laser with a wavelength of 509nmcombined with a semiconductor tapered amplifier (TA) [13,14]. However, the amplified power is still limited and the quality of the output beam is relatively poor, which limit its applications. Another method to use a high-power 1018nm fiber light source and then use frequency doubling to achieve high-power 509nm laser. This approach combined with laser frequency doubling through the seed source and fiber MOPA amplification can achieve a 509nm narrow linewidth laser with high power, good beam quality and excellent performance [15].

One of the challenges of laser frequency doubling method to achieve high power 509nm laser is to obtain high-power 1018nm single-frequency laser above the hundred-watt. So far, the output power of 1018nm lasers has reached the level of hundreds to even thousands of watts. For example, Tsinghua University achieved a 1018nm laser with a power of 1150W by using high-power oscillators in 2018[16]. However, this type of laser is generally used as a co-band pump light source [17]. Although other research institutions utilize high-power oscillators to achieve high-power 1018nm laser with MOPA amplification [18], the reported laser spectral are significantly wide, making it difficult to meet the demand for Rydberg atomic microwave measurement and other high-precision detection applications.

Generally, the commonly used 1018nm single-frequency seed sources for high power MOPA amplification are distributed feedback (DFB) or distributed Bragg-reflector (DBR) structures. The DFB fiber lasers are more difficult to manufacture because the inscription of phase-shifted gratings on doped fibers is required. In comparison, DBR fiber lasers can be fabricated by simply fiber fusion splicing, without the need to inscribe grating structures on rare-earth doped fibers. It is unnecessary to require the rare-earth-doped fiber to be photosensitive, which makes DBR fiber lasers more advantageous in terms of laser band, wavelength flexibility and so on. Therefore, The single-frequency fiber seed source with DBR structure can be a preferable choice for MOPA amplification at 1018nm.

The amplified spontaneous emission (ASE) is an important issue in high-power MOPA amplification at the 1018nm wavelength. Since the output power of DBR-type single-frequency narrow linewidth seed sources is generally in the order of 10mW, a gain of more than 40dB is required to amplify the output power to more than 100w. Therefore, multi-stage fiber amplification is a common solution, but ASE can easily be accumulated in this amplification process. At the same time, the 1018nm wavelength is close to the emission peak of Yb ions at 1030nm, which makes it extremely easy to induce self-oscillation of lasers near the 1030nm wavelength during the amplification process. Once the self-oscillation occurs, the gain of the fiber amplifier will shift to the 1030nm self-oscillation wavelength, which results the inability to further increase the output power of the 1018nm single-frequency laser. Another limitation to the power increase of the 1018nm single-frequency laser is stimulated Brillouin scattering (SBS), which is a common issue faced by all single-frequency fiber amplifiers. The usual solution is to choose gain and transmission fibers with shorter lengths or directly increase the mode field area of gain fibers. In conclusion, the utilization of appropriate fiber types and lengths to enhance ASE and SBS suppression and increase output power has become a challenge in obtaining high-power 1018nm laser.

This paper reports a 1018nm seed source with a linewidth less than 3k and an output power greater than 10mW by using a DBR structure. The wavelength tuning range of this seed source exceeds 1.5nm within the temperature range from 1°C to 95°C. It is observed that the seed source demonstrates strong linearity between the tuning wavelength and temperature, and no mode hopping occurred during the tuning process. The 1018nm laser with an output power exceeding 120W can be achieved by selecting appropriate doped fibers and optimizing their length. The experimental results show that the slope efficiency of the main amplification is greater than 77%, and the spectral ASE suppression ratio of the amplified laser is greater than 60dB. The linewidth and the beam quality factor $M^2$ of the amplified laser is about 10kHz and 1.3 respectively.

# 2 Experimental setup

2.1) Widely tunable single-frequency seed source at 1018nm

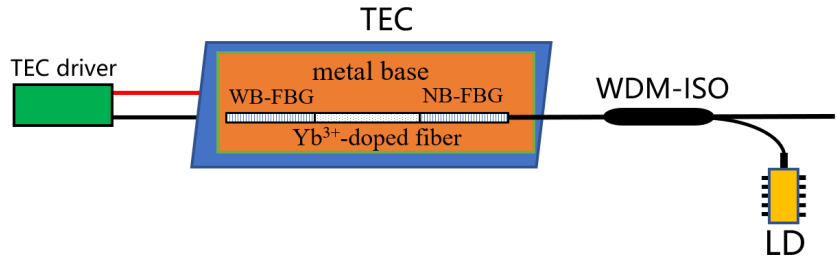

Fig. 1. Structure of single-frequency DBR fiber laser

The structure of single-frequency DBR fiber laser used in this experimental setup is depicted in Fig. 1. In this structure, a 974nm laser diode with single-mode fiber pigtail output is used as the pump source for the 1018nm seed laser. By backward pumping, the pump light first passes through an integrated 980/1018nm wavelength division multiplexer/isolator (WDM/ISO) device, and then is coupled into the DBR fiber laser.

And the resonant cavity consists of an 8mm-long segment of high-concentration $Yb^{3+}$-doped fiber (with a gain absorption coefficient exceeding 1200 dB/m at 977nm) and a pair of Bragg gratings. The center wavelength of both Bragg gratings is 1018 nm. The reflectivity of the wide-bandwidth fiber brag grating (WB-FBG) is greater than 99% with a reflection bandwidth of 0.20 nm; the narrow-bandwidth fiber brag grating (NB-FBG) has a reflectivity of 75% and a reflection bandwidth of 0.06 nm. Additionally, the WB-FBG has a length of 10 mm and is inscribed on a single-mode non-polarization-maintaining fiber, while the 20mm long NB-FBG is inscribed on a single-mode polarization-maintaining fiber. Besides, the entire resonant cavity is mounted on a metal base, which contains an embedded Thermo Electric Cooler (TEC). Temperature control of this resonant cavity within a range of 0 to 100°C is achieved by driving the TEC through a control circuit board.

2.2) The experimental setup for high-power amplification with narrow 1018 nm seed

The 1018nm narrow linewidth all-fiber amplifier is shown in Fig. 2. The signal light from the self-made DBR-type 1018nm seed source, which first passes through an isolator ISO1, then couples into a polarization-maintaining single-mode doped fiber 1 together with the single-mode pump light from LD1. The corresponding wavelength of the LD1 pump light is 974 nm, with a maximum output power of 400 mW. After the single-mode amplification, the single-frequency laser first passes through a filter 1 with a bandwidth of 2nm at 1018nm and an isolator ISO2. Subsequently, it passes through a coupler 1 combined with the pump light from LD2, and finally couples into a 10/125 polarization-maintaining doped fiber for second-stage amplification. The LD2 is a 976nm wavelength-locked multi-mode pump source with a maximum output power of 9W.

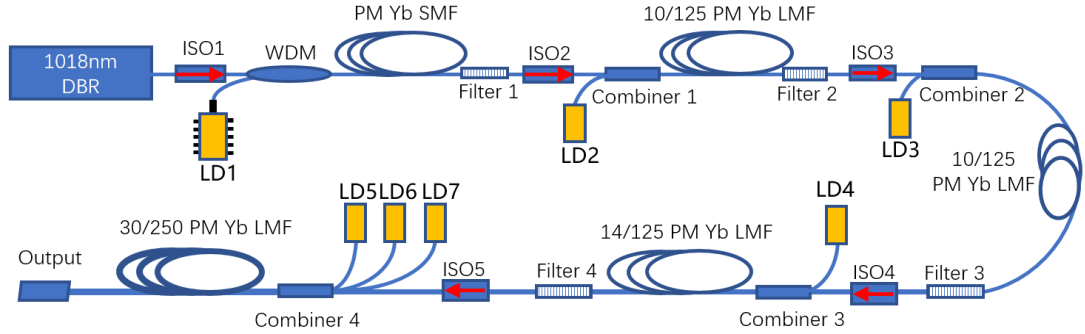

Fig. 2. Schematic diagram of a single-frequency optical fiber amplifier at 1018 nm

After the second-stage amplification, the laser first passes through a filter 2 with a bandwidth of 2nm at 1018nm and an isolator ISO3. Subsequently, it passes through a coupler 2 combined with the pump light from LD3, and finally couples into a 10/125 polarization-maintaining doped fiber for third-stage amplification. The LD3 is a 976nm wavelength-locked multi-mode pump source with a maximum output power of 9W. After the third-stage amplification, the laser first passes through a filter 3 with a bandwidth of 2nm at 1018nm and an isolator ISO4. Then it passes through a coupler 3 combined with the pump light from LD4, and couples into a 14/125 polarization-maintaining doped fiber for fourth-stage amplification. The LD4 is a 976nm wavelength-locked multi-mode pump source with a maximum output power of 27W. After the fourth-stage amplification, the laser first passes through an isolator ISO5 and a filter 4, which is identical to filters 1, 2, and 3. Then it passes through a coupler 4 combined with the pump light from LD5, LD6, and LD7, and finally couples into a 30/250 polarization-maintaining doped fiber for fifth-stage amplification. LD5, LD6, and LD7 are also 976nm wavelength-locked multimode pump sources, with a maximum output power of 60W.

## 3 Experimental results and analysis

After the fabrication of the single-frequency DBR fiber laser seed source, its longitudinal mode state, wavelength, and linewidth parameters were tested.

In order to test the longitudinal mode performance of the seed laser as shown in Figure 1, the pump light power from LD is set to 300 mW, and the temperature controller of the DBR resonant cavity is set to 25 °C. Then the output 1018nm signal light with a power of approximately 10mW from the seed source is transmitted to a longitudinal mode detection system, which is composed of a scanning Fabry-Perot interferometer (Thorlabs, SA210), a signal generator (Thorlabs, SA201), and an oscilloscope (Tektronix, DPO2024). The resolution and Free Spectral Range (FSR) of the F-P interferometer in this detection system are 67 MHz and 10 GHz respectively. From the test result shown in Figure 3, it can be confirmed that the seed laser is stabilized in only one single longitudinal mode.

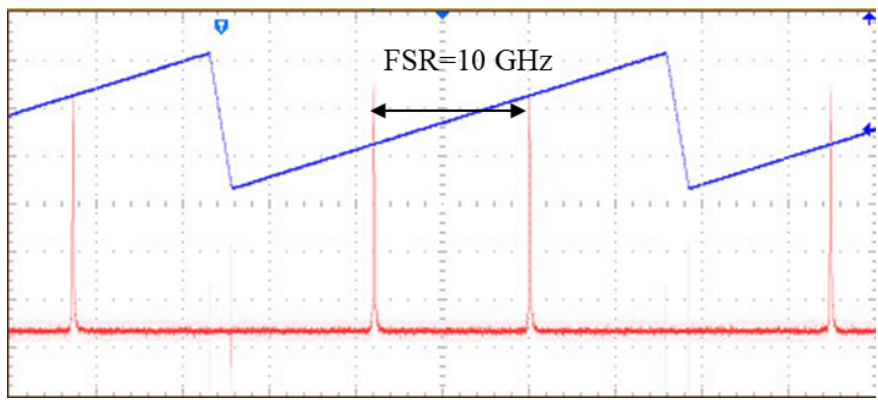

Fig. 3. The single longitudinal mode test of the laser seed source

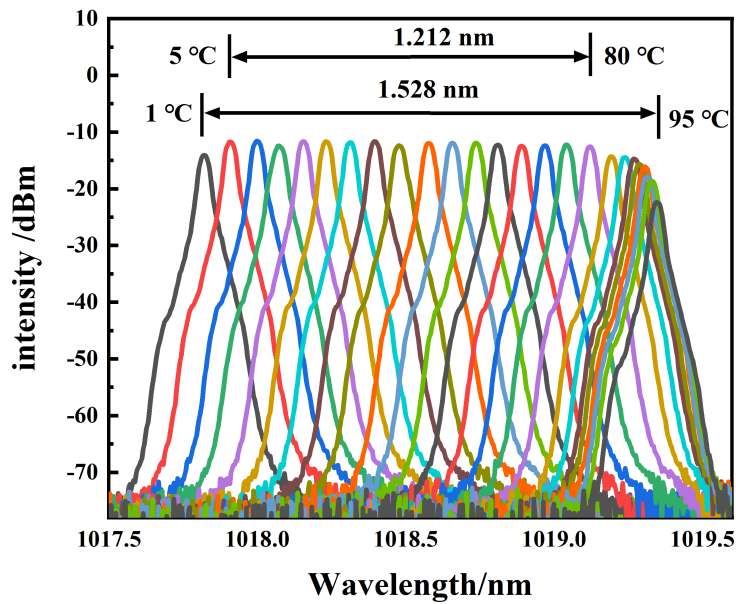

Fig. 4. Spectra of the output laser at different temperatures

In order to test the wavelength tuning performance of the seed laser, the temperature of the DBR resonant cavity is gradually changed from 1°C to 95°C through a TEC temperature control circuit system, and the corresponding tuning wavelength range of the laser is showed in Fig. 4. It can be see that the center wavelength of the laser correspondingly shifts from 1017.83 nm to 1019.36 nm as the temperature of the resonant cavity increases from 1°C to 95°C. The tuning range of this seed laser is greater than 1.5 nm, and there is no significant attenuation in the output power as the tuning range spans 1.212 nm while the temperature changes from 5°C to 85°C. The Fabry-Perot interferometer is monitoring the longitudinal modes throughout the entire wavelength tuning process. And the monitoring result indicates the output laser consistently maintains a single longitudinal mode state without mode hopping. The evolution of the wavelength versus temperature of the DBR resonant cavity gives us a wavelength map as Figure 5, from 1°C to 95°C. It can be observed that the wavelength varies linearly with tuning temperature, and the linearity coefficient R is 0.99972, which indicates that the DBR seed source produced in this article has good linear temperature tuning performance.

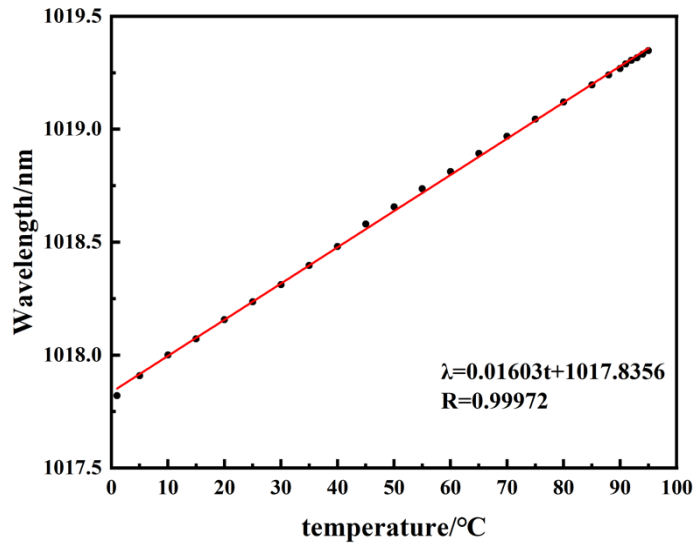

Fig. 5. Wavelength variation of the output laser with temperature

A multi-stage amplification experimental optical path is constructed according to the schematic diagram 2. The output power of the self-made seed source after passing through an isolator is approximately 10 mW. After the first-stage single-mode polarization-maintaining amplification and spectral filtering, the amplified output power is approximately 170 mW when the pump power is increased to 400mW. In the second-stage amplification after passing through an isolator, the length of the doped fiber is trimmed to less than 2 m to ensure the output power and suppress self-oscillation near the 1030nm spectra. When the pump power is increased to 6W, the output power is greater than 1.7W after passing through a spectral filter and an isolator. In the third-stage polarization-maintaining amplification, the length of the 10/125 polarization-maintaining doped fiber is further optimized. When the pump power is increased to 8W, the output power exceeds 4.5W after passing through a filter and an isolator. To suppress self-excited oscillations, the length of the doped fiber in the fourth-stage amplification is optimized to approximately 1.5 meters. When the pump power is increased to 20W, the output power finally exceeds 14W after passing through a filter and an isolator.

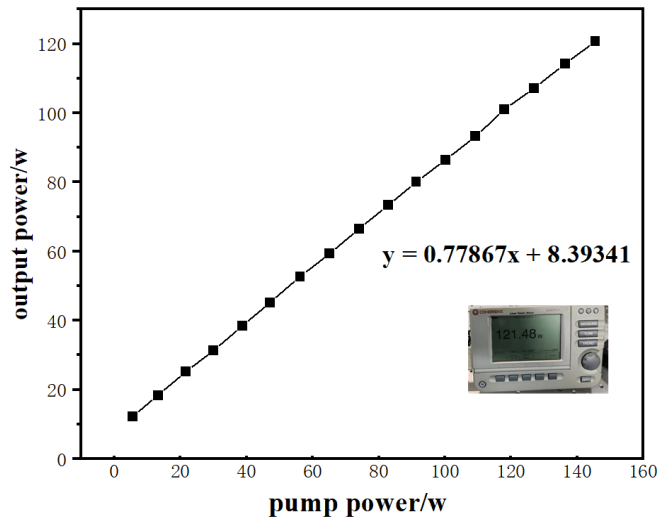

Fig. 6. Optical-to-optical conversion efficiency of the main amplified light at 1018nm

The optical power of the amplified signal light increases linearly with the pump power as depicted in Figure 6, which indicates the amplifier is clearly unsaturated. When the pump power is increased to 145W, the output power exceeds 120W, and the slope efficiency of this main amplifier is 77.8%.

In the experiment, it was found that there are three situations where self-excited oscillations are prone to occur. Firstly, if the laser output power exceeds 1.7W in the second-stage amplification, further increasing the pump power can easily result in self-oscillation. Secondly, if the output light with a power of 1.7W in the second-stage amplification is directly coupled into the 14/125 doped fiber for amplification, self-oscillation can easily occur when the output power exceeds 10W. Finally, if the 1018nm signal light is coupled into the main amplifier at a power below 10W, self-oscillation can also easily occur when the output power exceeds 80W. And its signal-to-noise ratio of the spectrum will be comparatively low.

The spectrum of 1018nm main amplification is sketched in Figure 7. As shown in the figure, although there is still an ASE background noise at 1030nm without filtering, the peak of the 1018nm signal light is approximately 60dB higher than the ASE. The peak of the residual pump light near 976nm is more than 40dB lower than the main peak at 1018nm. In summary, the output laser mainly consists of 1018nm signal light and has a good signal-to-noise ratio.

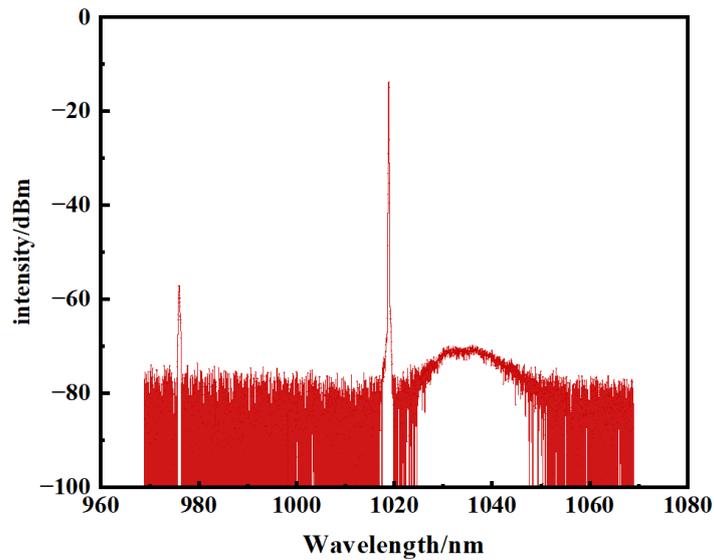

Fig. 7. Spectra of the output laser from the main amplification

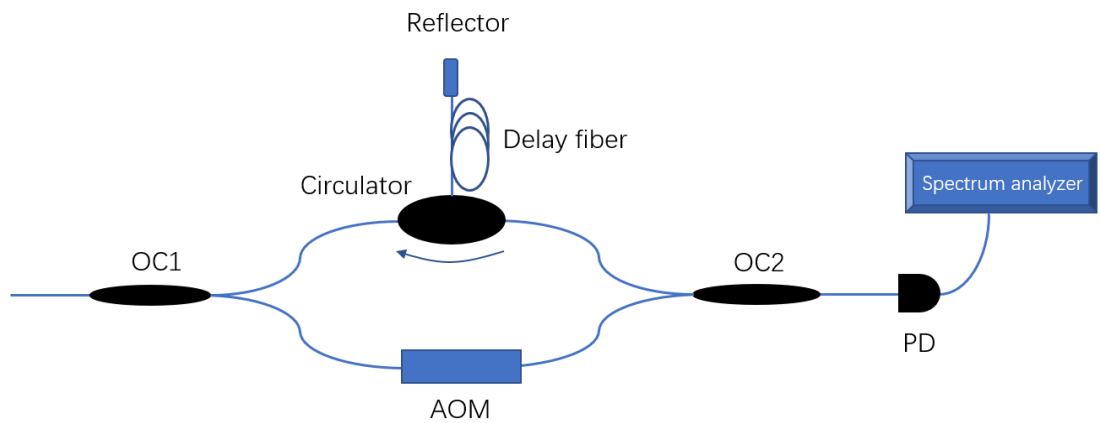

Fig. 8. Structure of the linewidth detection system

In order to measure the linewidth of output light, a linewidth detection system based on the principle of self-heterodyne beat frequency has been constructed as illustrated in Figure 8. It can be seen that the components of this detection system mainly include a coupler, circulator, 18 km-long delay fiber, 1018 nm reflector, acousto-optic modulator, photodetector, and electronic spectrum analyzer. A DFB seed source (NKT Photonics, Koheras Adjustik Y10) is used to validate the measurement effectiveness of this linewidth detection system. Figure 9 displays that the width at 20 dB below the peak of the beat signal is 77 kHz, which should be approximately 20 times wider than the actual linewidth[12]. Hence, the measured linewidth of this DFB seed is 3.85 kHz, which matches its nominal linewidth. It can be concluded that the detection system is effective.

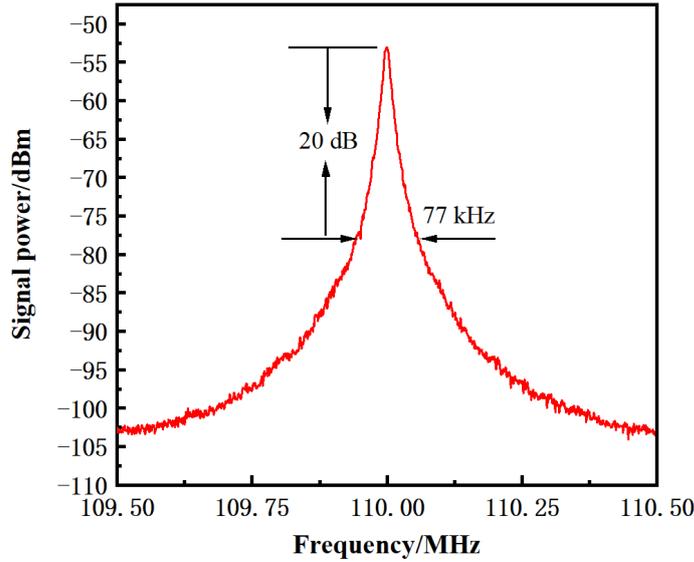

Fig. 9. Measured linewidth of a DBR fiber laser (NKT Photonics, Koheras Adjustik Y10)

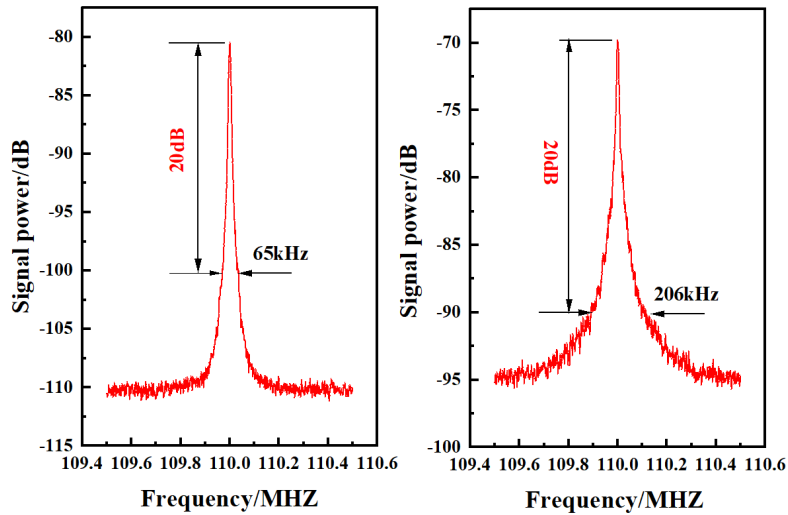

Fig. 10. Measured linewidth of seed laser and amplified laser

The left plot is about seed laser, and the right plot is about amplified laser.

To measure the output laser linewidth of the main amplifier, a beam splitting and spatial coupling optical system is added at the output end of the 120W amplified spatial laser. A small portion of 120W output light is recoupled into a single-mode fiber and the linewidth detection system. Figure 10 displays the test results obtained by separately measuring the linewidths of the DBR laser seed and the amplifier. It can be seen that the full width at half maximum (FWHM) at 20 dB below the peak of the beat signal is 65 kHz in the seed source linewidth measurement, and the linewidth of the seed source is about 3.25 kHz. Similarly, the full width at half maximum (FWHM) of the amplified output light is approximately 206 kHz, and its actual linewidth is 10.3 kHz, which is 3.17 times wider than that of the seed source. The broadening of linewidth is probably attributed to the

ASE noise generated during the amplification process and the pump noise from pump LDs.

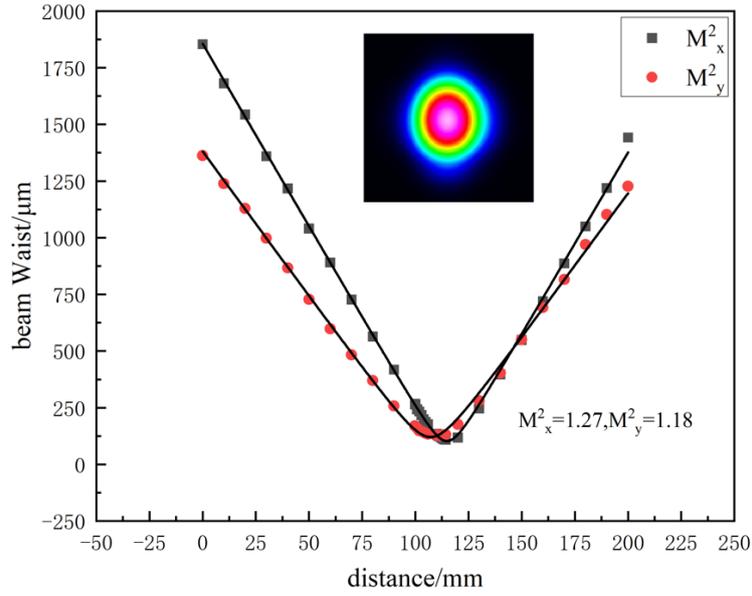

Fig. 11. Measured beam quality factor $M^2$ of amplified laser

In order to measure the output beam quality, the amplified 1018nm laser passes through a polarized beam splitter (PBS) and is attenuated before being coupled into a beam quality measurement instrument (Thorlabs, BP209-IR/M). Figure 11 shows the measurement results. It can be seen from the figure that the quality of the main amplified output laser beam is $M^2_x=1.27$ and $M^2_y=1.18$, indicating excellent output laser performance. Besides, the measured polarization degree is around 19dB when the amplified laser is coupled into another identical experimental optical path.

## 4 Conclusion

In this paper, a 1018nm seed source with a linewidth less than 3kHz is achieved by using a DBR structure, and an output power greater than 10mW. The wavelength tuning range of this seed source exceeds 1.5nm within the temperature range from 1°C to 95°C. It is observed that the seed source demonstrates strong linearity between the tuning wavelength and temperature, and no mode hopping occurred during the tuning process. The 1018nm laser with an output power exceeding 120W can be achieved by selecting appropriate doped fibers and optimizing their length. The experimental results show that the optical-to-optical conversion efficiency of the main amplification is greater than 77%, and the spectral ASE suppression ratio of the amplified laser is greater than 60dB. The linewidth and the beam quality factor $M^2$ of the amplified laser is about 10kHz and 1.3 respectively. In addition to the length and mode field area of the doped fiber, the power ratio between input and output of the main amplifier also has an important effect on ASE suppression.